\documentclass{svproc}
\usepackage{url}

\usepackage{graphicx}
\usepackage{multicol}
\usepackage{footmisc}
\usepackage{amsfonts}
\usepackage{mathtools}

\usepackage{xcolor}
\usepackage{subcaption}
\usepackage[ruled]{algorithm2e}
\usepackage{dsfont}

\usepackage[commentmarkup=todo]{changes}
\definechangesauthor[color=red]{is}

\DeclareMathOperator*{\argmax}{argmax}

\usepackage{hyperref}

\begin{document}

\mainmatter              
%
\title{ELRUHNA: Elimination Rule-based Hypergraph Alignment} 
\titlerunning{Hypergraph Alignment}  
%
\author{Cameron Ibrahim\inst{1} \and S M Ferdous\inst{2}
 Ilya Safro \inst{1} \and Marco Minutoli\inst{2} \and Mahantesh Halappanavar\inst{2}}
\authorrunning{Cameron Ibrahim et al.} 
%
\tocauthor{Cameron Ibrahim, S M Ferdous, Ilya Safro, Marco Minutoli, and Mahantesh Halappanavar}
\institute{University of Delaware, Newark, DE 19716, USA,\\
\email{\{cibrahim,isafro\}@udel.edu},\\ 
\and
Pacific Northwest National Laboratory,
Richland WA 99354, USA, \\
\email{\{sm.ferdous, marco.minutoli, hala\}@pnnl.gov}
}

\maketitle              

\begin{abstract} 
Hypergraph alignment is a well-known NP-hard problem with numerous practical applications across domains such as bioinformatics, social network analysis, and computer vision. Despite its computational complexity, practical and scalable solutions are urgently needed to enable pattern discovery and entity correspondence in high-order relational data. The problem remains understudied in contrast to its graph based counterpart. In this paper, we propose ELRUHNA, an elimination rule-based framework for unsupervised hypergraph alignment that operates on the bipartite representation of hypergraphs. We introduce the incidence alignment formulation, a binary quadratic optimization approach that jointly aligns vertices and hyperedges. ELRUHNA employs a novel similarity propagation scheme using local matching and cooling rules, supported by an initialization strategy based on generalized eigenvector centrality for incidence matrices. Through extensive experiments on real-world datasets, we demonstrate that ELRUHNA achieves higher alignment accuracy compared to state-of-the-art algorithms, while scaling effectively to large hypergraphs. 

\noindent {\bf Reproducibility:} Code and data are available at [link will be added upon acceptance].
\end{abstract}


\section{Introduction}

Given hypergraphs \(H_Q\coloneqq (V_Q, E_Q)\) and \(H_D \coloneqq (V_D, E_D)\) known as the query hypergraph and data hypergraph respectively, the \emph{hypergraph alignment} problem aims to find an injection \(\sigma\colon V_Q \hookrightarrow V_D\) maximizing some measure of  overlap between $H_Q$ and $H_D$. Typical applications of the hypergraph alignment include finding patterns common between a pair of hypergraphs, or identifying vertices which represent the same entity in the query and data hypergraphs. This problem extends the \emph{graph alignment} with the goal to capture higher-order relationships between nodes. Hypergraph alignment is of  considerable interest in fields such as  machine learning \cite{doUnsupervisedAlignmentHypergraphs2024}, bioinformatics \cite{aladaug2013spinal,neyshabur2013netal}, computer vision \cite{HNNHM}, and natural language processing \cite{mao-etal-2021-alignment}. 

For example, in machine learning and computer vision, the hypergraph alignment methods are commonly used to perform feature matching \cite{doUnsupervisedAlignmentHypergraphs2024}. Given two datasets, we attempt to represent a set of features as the vertices of a  hypergraph. Hypergraph alignment can then be used as a heuristic method for identifying features which are common between the two datasets. Another common framing of this problem in machine learning is to attempt to identify user correspondence across social networks\cite{user_correspondence}. In this setting, these social networks are represented as hypergraphs with users as vertices,  groups of friends/co-authors/coworkers/etc as hyperedges. As a result of privacy concerns, methods in this area may consist of supervised and unsupervised approaches \cite{doUnsupervisedAlignmentHypergraphs2024}.  


In this paper, we focus on finding global solutions to the hypergraph alignment problem in an unsupervised setting. In this context, the goal is to find a mapping \(\sigma\) which maximizes the hyperedge correctness metric. In short, \(\sigma\) is chosen to maximize the number of hyperedges \(e\) in the query hypergraph such that \(\{\sigma(v) \mid v \in e \}\) is also an hyperedge in the data hypergraph \cite{HNNHM}. 
Maximizing hyperedge correctness for general (hyper)graphs is NP-hard \cite{HNNHM}. 

\noindent \textbf{Our Contribution:} We  introduce a number of variants on the hyperedge correctness metric for  hypergraph alignment which we refer to collectively as \emph{incidence alignment} (Eq.\ \ref{eq:incidence}) and discuss the benefits of this formulation. We provide and efficient heuristic for maximizing incidence alignment (Algorithm \ref{alg:elruhna}) and introduce a novel hypergraph alignment solver ELRUHNA that exhibits superior quality-runtime trade-off. We compare the results obtained by ELRUHNA against state of the art methods, and demonstrate an improvement in quality of up to 25\%, and solve problem instances with tens of thousands of vertices in the bipartite representation.

\section{Background \& Notation}

An undirected graph \(G = (V, E)\) consists of a vertex set \(V\) and an edge set \(E \subseteq \{e \in 2^V \mid \lvert e \rvert = 2\}\). The higher order generalization of an undirected graph is a hypergraph \(H = (V, E),\) whose set of hyperedges is a subset \(E \subseteq 2^V\). The rank of a hypergraph is the maximum size of any of its edges. 
A vertex \(v\) is incident to a hyperedge \(e\) if \(v \in e.\) Two vertices \(v,u\) are adjacent if there exists a hyperedge \(e\) which is incident to both \(u\) and \(v.\) A graph \(G = (V,E)\) is bipartite if \(V = V_1 \cup V_2\) such that $V_1\cap V_2 = \emptyset$ and \(uv \in E\) implies \(u \in V_1,v\in V_2\) or vice versa.  \(V_1\) and \(V_2\) may be referred to as the left and right parts respectively. The bipartite representation of a hypergraph $H=(V,E)$ is a bipartite graph \(G^\prime = (V \cup E, E^\prime)\) such that the left and right parts are sets $V$ and $E$, respectively, and a vertex \(v\in V\) is adjacent to a vertex that represents a hyperedge \(e \in E\) in \(G^\prime\) if and only if $v\in e$ in \(H\), i.e., there is an edge $ve\in E'$. The clique expansion of a hypergraph \(H=(V,E)\) is an undirected graph \(G^\prime \coloneqq (V, E^\prime)\) defined so that \(u\) and \(v\) are adjacent in \(G^\prime\) if and only if \(u\) and \(v\) are adjacent in \(H\), i.e., each hyperedge in a hypergraph is represented as a clique (that contains all its nodes) in the graph. 

Throughout this paper, we make use of the indicator function \(\mathds{1}[\phi] \coloneqq 1\) if \(\phi\) is true and 0 otherwise. 
For a hypergraph \(H=(V,E)\) with \(n \coloneqq \lvert V\rvert\) vertices and \(m \coloneqq \lvert E\rvert\) hyperedges, we can define an \(n\times n\) matrix \(A\) and an \(n \times m\) matrix \(B\) as \(A_{uv} \coloneqq \mathds{1}[uv \in E]\) and \(B_{ue} \coloneqq \mathds{1}[u \in e, ~ e \in E].\)
\(A\) and \(B\) are known as the adjacency and  incidence matrices of \(H\), respectively. Notably, \(A\) is also the adjacency matrix of the clique expansion of \(H\), and so elides any higher order information. 
Weighted variants \(\tilde A, \tilde B\) of these matrices may be created by replacing any \(1\) with a positive real number.

\subsection{Hypergraph Alignment} 

Broadly, hypergraph alignment aims to map a given query hypergraph \(H_Q \coloneqq (V_Q, E_Q),\) onto a known data hypergraph \(H_D\coloneqq (V_D, E_D).\) To do so, we attempt to find an injective function \(\sigma\colon V_Q \hookrightarrow V_D\) which maximizes some reward function \(R\)
\begin{align*}
    \max_{\sigma\colon V_D \hookrightarrow V_Q} R(H_D, H_Q, \sigma).
\end{align*}

If the ground truth \(\sigma^*\) is known for some subset \(S \subseteq V_Q,\) we may attempt to maximize accuracy
\(
    R_{acc}(\sigma) = \frac{1}{\lvert S\rvert}\sum_{v \in S} \mathds{1}\left[\sigma(v) = \sigma^*(v)\right].
\)
In this paper, we focus on the unsupervised case where the ground truth \(\sigma^*\) is not known during the optimization process \cite{doUnsupervisedAlignmentHypergraphs2024,chenCONEAlignConsistentNetwork2020}.

A common metric used in  both hypergraph and graph alignments literature is the hyperedge correctness
\begin{align*}
    R_{EC}(\sigma)  &= \frac{1}{\lvert E_Q\rvert} \sum_{e \in E_Q} \mathds{1}\left[\{\sigma(v) \mid v \in e\} \in E_D\right].
\end{align*}
This gives the fraction of hyperedges in the query hypergraph which are present in the data hypergraph after mapping. 

A mapping \(\sigma\colon V_Q \hookrightarrow V_D\) can alternatively be represented by an \(\lvert V_Q\rvert \times \lvert V_D\rvert\) binary matrix \(X\) whose rows and columns all sum to at most one \cite{bayati_sparse}. This can be expressed succinctly in terms of the \(\lVert X\rVert_1\) and \(\lVert X\rVert_\infty\) norms, defined as 
\(
    \lVert X \rVert_1 = \max_{1 \leq j\leq n}  \sum_{i=1}^{m}\lvert X_{ij}\rvert\) and \( \lVert X \rVert_{\infty} = \max_{1 \leq i \leq m} \sum_{j = 1}^{n} \lvert X_{ij}\rvert.
\) respectively.
As such, hyperedge correctness can also be generalized as the following binary optimization problem. 
Let \(\tens{S}^{(k)}\) be a \(k\) dimensional tensor defined as  \[\tens{S}^{(k)}_{(v_1, u_1), \ldots, (v_k, u_k)} = \begin{cases}
    1 & \{v_1,\ldots, v_k\} \in E_D \wedge \{u_1, \ldots, u_k\} \in E_Q,\\
    0 & \text{otherwise}.
\end{cases}\]
If \(K\) is the rank of the hypergraph, the  hypergraph alignment problem is then generalized by
\begin{align*}
\max_{X}\qquad &\langle W, X\rangle_F + \beta \sum^{K}_k \tens{S}^{(k)}\text{vec}(X)^{\otimes k} \\
\text{st}\qquad & 
    \lVert X \rVert_1 ,\lVert X \rVert_\infty \leq 1\\
    & X \odot \Pi = X\\ 
& X \in \{0, 1\}^{\lvert V_Q\rvert \times \lvert V_D\rvert},
\end{align*}
where \(\langle W, X\rangle_F = \text{Tr}[W^TX]\) is the Frobenius inner product and \(X \odot \Pi = X\) is a constraint which can prohibit certain pairs from being matched. Setting \(W = 0\) and \(\beta = \frac{1}{\lvert E_Q\rvert}\) gives a result which maximizes \(R_{EC}\).

For a general hypergraph with \(K > 2\), this is an example of the NP-hard polynomial integer programming problem \cite{friedlandSpecialIssuePolynomial2022a}.
Using the equivalence \(\tens{S}^{(2)} = A_Q \otimes A_D\) where \(\otimes\) is the Kronecker product,  hypergraph alignment with \(K=2\) is known as graph alignment and is an example of integer quadratic programming, for which there are a variety of successful heuristics \cite{xiangCuAlignScalableNetwork2023,bayati_sparse,ShortSurveyRecent}. 
\begin{align}\label{eq:network_align}
    \begin{split}
           \max_{X}\quad & \langle W, X\rangle_F + \frac{\beta}{2}\langle XA_Q, A_DX \rangle_F\\
    \text{st}\quad &  
    \lVert X \rVert_1 ,\lVert X \rVert_\infty \leq 1\\
    & X \odot \Pi = X\\ 
    & X \in \{0, 1\}^{\lvert V_Q\rvert \times \lvert V_D\rvert} 
    \end{split}
\end{align}


Furthermore, when the data and query graph have an equal number of vertices there exists a constant \(C\) not dependent on \(X\) such that the quadratic portion of the objective is equivalent to \(
    \langle XA_Q, A_DX \rangle_F + C = \frac{1}{2}\lVert X^TA_QX - A_D\rVert_F^2 \)
which is a common way of framing the edge correctness objective for graph alignment \cite{bigalign}.

\subsection{Related Work} \label{sec:related}

Due to its range of applications, graph alignment has been subject to a broad array of approaches. While graph alignment has been extensively studied, hypergraph alignment remains relatively underexplored. In this section, we will provide an overview of relevant research in both problems.
Further information regarding these methods can be found in \cite{ShortSurveyRecent}.

A variant of graph alignment for bipartite graphs has also been discussed in \cite{bigalign} with the algorithm BiG-Align. In this context, both \(G_Q\) and \(G_D\) are bipartite, and the aim is to match the left (resp. right) part of \(G_Q\)  to the left (resp. right) part of \(G_D\). To do so, two matching matrices \(X, Y\) are found to maximize \(\lVert X^TB_QY - B_D\rVert^2_F.\) 
In \cite{doUnsupervisedAlignmentHypergraphs2024}, the authors have made use of this formulation using the equivalency of hypergraphs to bipartite graphs in order to create a heuristic for hypergraph alignment. We utilize a similar formulation in our method.

 One may attempt to solve a graph or hypergraph alignment problem by directly trying to solve the integer programming formulations; however, because of the computational cost of solving these problems with conventional solvers, a number of approximate methods have arisen. Big-Align is one of several methods which attempt to solve the graph alignment problem by relaxing the integer constraint on the optimization problem. Doing so produces a matrix which may be interpreted as a set of probabilities indicating how likely it is to match pair of vertices. Approximate matching algorithms such as those described in \cite{xiangCuAlignScalableNetwork2023,mahantesh_multithread_matching} may then be used in order to round these to a solution. In NetAlign, a belief propogation algorithm is applied to find an approximate solution to the integer quadratic programming problem of graph alignment \cite{AlgorithmsLargeSparse}.  In Adaptive Discrete Hypergraph Mathching, the integer polynomial programming representation of the hypergraph matching problem is solved via iterative linear approximations \cite{yanAdaptiveDiscreteHypergraph2018}.

Of particular importance to this paper is the ELRUNA algorithm \cite{qiuELRUNAEliminationRulebased2021}. ELRUNA is an algorithm for graph alignment based purely on the topology of the input graphs. In ELRUNA, a set of initial similarities between vertices in the query and data graphs are iteratively smoothed according to a number of elimination rules. These rules govern which similarities are allowed to propagate through the graph by tying each update to the result of a local matching problem in the region of a pair. An approximation algorithm is then again used in the matrix of similarities in order to achieve a result.

Node embeddings are commonly used in (hyper)graph alignment to determine similarities between pairs of nodes and determine candidate pairs for the final alignment. The Cone-align algorithm embeds each graph separately, the attempts to align each embedding space in order to find a unified embedding for both graphs \cite{chenCONEAlignConsistentNetwork2020}. This technique is extended by the cuAlign algorithm, which includes sparsification of candidate pairs to reduce computational complexity, then solves the resulting sparsified problem using a combination of belief propagation and an approximate weighted matching solver \cite{xiangCuAlignScalableNetwork2023}.

Node embeddings have also been seen in a hypergraph context with approaches such as HyperAlign \cite{doUnsupervisedAlignmentHypergraphs2024} and HNN-HM \cite{liaoHypergraphNeuralNetworks2021}. In the former, features for each node are extracted from each hypergraph, then the resulting data is embedded using contrastive learning. The node emeddings are then used to determine similarities between nodes, which are utilized to sparsify the inputs as in cuAlign. In HNN-HM, the data and query hypergraphs are combined into a single association hypergraph, which is then embedded as a single object before being decoded and classified.

\section{Algorithm ELRUHNA: Elimination Rule-based Hypergraph Network Alignment}

In this section, we will introduce two variations of the hypergraph  alignment optimization problem which we refer to as \emph{incidence alignment}. The aim of incidence alignment is not only to align the vertices of the data and query hypergraphs, but to align the hyperedges as well in order to maintain the incidence of the aligned hypergraphs as much as possible. 

After this, we will introduce the Elimination Rule-based Hypergraph Network Alignment algorithm, or ELRUHNA. The full algorithm is presented in Algorithm \ref{alg:elruhna}. ELRUHNA utilizes an iterative strategy to propagate similarities throughout the hypergraph using a pair of elimination rules which are described in Sect. \ref{sec:elimination_rules}. Strategies for determining initial similarities are discussed in Sect. \ref{sec:initialization}.

\begin{algorithm}\caption{Elimination Rule Based Hypergraph Alignment}\label{alg:elruhna}
    \textbf{Input:} Incidence matrices \(B_Q, B_D\) with sizes \(n_Q, m_Q\) and \(n_D, m_D\)\\
    \textbf{Output:} Matrices \(X,Y\) representing \(\sigma_V, \sigma_E\)\\
    \(X, Y \leftarrow \texttt{zeros}(n_Q, n_D), \texttt{zeros}(m_Q, m_D)\)\;
    \While{\(\texttt{has\_unmatched}(X, Y)\)}{
        \(X^*, Y^* \leftarrow \texttt{matched}(X, Y)\)\;
        \(W_V, W_E \leftarrow \tilde B_Q^T Y^*\tilde B_D, \tilde B_QX^*\tilde B_D^T\),  Equation \ref{eq:partial_solution}\;
        \(B_Q^\prime, B_D^\prime, W_V^\prime, W_E^\prime \leftarrow \texttt{get\_unmatched}(B_Q, B_D, W_V, W_E, X^*, Y^*)\)\;
        $X \leftarrow \texttt{initialize\_similarity}(\tilde B_Q^\prime, \tilde B_D^\prime, W_V^\prime, W_E^\prime)$, Sect. \ref{sec:initialization}\;
        \(Y \leftarrow W_E^\prime + \tilde B_Q^\prime X^\prime(\tilde B_D^\prime)^T\)\;
        Propagate similarities, Algorithm \ref{alg:propagate}\;
        $Y^\prime \leftarrow \texttt{dominant\_match}(Y^\prime)$\;
        \(X^\prime \leftarrow \texttt{dominant\_match}(W_V + (\tilde B_Q^\prime)^T Y^\prime\tilde B_D^\prime)\)\;
        $X, Y \leftarrow \texttt{update}(X, Y, X^\prime, Y^\prime)$\;
    }
    
\end{algorithm}

\subsection{Incidence Alignment}

In practical contexts, the goal of hypergraph alignment is often to try to identify patterns shared across the data and query instances, $H_Q$ and $H_D$ in order to identify vertices representing entities present in both hypergraphs. In real world scenarios, these patterns may be prone to noise: for example, an individual may be a member of a friend group in a social network represented by $H_Q$, but not in the network represented by $H_D$. As a result, if one were to consider the hyperedge \(e\) representing this friend group, it would be impossible to find a \(\sigma\) such that mapping \(e\) would contribute to the hyperedge correctness objective.

Instead, it may be more sensible to maximize the greatest overlap a hyperedge in $H_Q$ has with that in $H_D$. To do so, we simultaneously try to find an alignment \(\sigma_V\) of the vertices of \(H_Q\) to those of \(H_D\) and an alignment \(\sigma_E\) of the hyperedges of $H_Q$ to those of $H_D$. This is represented by the hyperedge overlap objective  
\begin{align}\label{eq:overlap}
    R_{O}(\sigma_V, \sigma_E) &\coloneqq  \sum_{e \in E_Q}\lvert \{\sigma_V(v) \mid v \in e\} \cap \sigma_E(e) \rvert.
\end{align}
The hyperedge overlap objective as presented here is then resistant to small differences between a hyperedge in \(H_Q\) and some corresponding ground truth hyperedge in \(H_D.\) 

Expanding out the intersection terms, we see that hyperedge overlap objective is equivalent to 
\begin{align*}
    R_{O}(\sigma_V, \sigma_E) = \sum_{e \in E_Q} \sum_{v \in V_Q} \mathds{1}[v \in e \wedge \sigma_V(v) \in \sigma_E(e) ].
\end{align*}
As such, another interpretation of the hyperedge overlap objective is that we are attempting to align the vertices and hyperedges of $H_Q$ and $H_D$ in order to maximize the number of incident pairs \((u,e)\) which are present in $H_D$ as \((\sigma(u), \sigma(e)).\) 

To improve our setup further, note that \(\lvert \{\sigma_V(v) \mid v \in e\} \cap \sigma_E(e) \rvert\) has a value of at most \(\min(\lvert e \rvert,  \sigma_E(e)).\) As a result, \(\{\sigma_V(v) \mid v \in e\}  = \sigma_E(e)\) contributes just as much to the objective as \(\{\sigma_V(v) \mid v \in e\}  \subset \sigma_E(e)\). In order to encourage matching hyperedges of similar cardinalates, we introduce the hyperedge incidence objective
\begin{align*}
    R_{I}(\sigma_V, \sigma_E) &= \frac{1}{\lvert E_Q\rvert} \sum_{e \in E_Q} \underbrace{\frac{\lvert \{\sigma_V(v) \mid v \in e\} \cap \sigma_E(e) \rvert}{\sqrt{\lvert e\rvert\lvert \sigma_E(e) \rvert}}}_{\text{overlap term}}.
\end{align*}
Each overlap term
is equal to one if and only if \(e\) aligns perfectly to \(\sigma(e).\) As a result, for any \(\sigma_V\) there exists \(\sigma_E\) such that \(R_{EC}(\sigma_V) \leq R_{I}(\sigma_V, \sigma_E)\) and \(R_I(\sigma_V, \sigma_E)=1\) if and only if \(R_{EC}(\sigma_V)=1.\)

Given an incidence matrix \(B\) of a hypergraph \(H,\) let \(\tilde B\) be the hyperedge normalized incidence matrix of \(H,\) defined as \(\tilde B_{ue} = \frac{B_{ue}}{ \sqrt{\lvert e \rvert}}\). If we utilize \(X,Y\) as the matrix representations of \(\sigma_V\) and \(\sigma_E\) respectively, we can rewrite this objective as \(\langle \tilde B_Q Y , X\tilde B_D\rangle_F.\) 
As a result, we may define the more general incidence alignment problem as
\begin{align}\label{eq:incidence}
    \begin{split}
    \max_{X,Y}\quad & \langle W_V, X\rangle_F + \langle W_E, Y\rangle_F + \beta\langle \tilde B_Q Y ,  X\tilde B_D \rangle_F\\
    \text{such that}\quad & \begin{aligned}
    & \lVert X \rVert_1 ,\lVert X \rVert_\infty \leq 1, &&  \lVert Y \rVert_1 ,\lVert Y \rVert_\infty \leq 1
    \\
    & X \odot \Pi_V = X, &&  Y \odot \Pi_E = Y\\ 
    &X \in \{0, 1\}^{\lvert V_Q\rvert \times \lvert V_D\rvert}, & &Y \in \{0, 1\}^{\lvert E_Q\rvert \times \lvert E_D\rvert}
\end{aligned}
\end{split}
\end{align}

Notably, this formulation for \(k=2\) is a quadratic program rather than a polynomial program. Furthermore, for \(W_V = W_E = 0,\) \(\lvert V_Q\rvert \leq \lvert V_D\rvert,\) and \(\lvert E_Q\rvert \leq \lvert E_D\rvert\),  Eq.\ \ref{eq:incidence} functions as a weighted extension of the objective used by BiG-Align as described in Sect. \ref{sec:related}.

It may be the case that we want to allow multiple hyperedges in the query hypergraph to be mapped to the same hyperedge in the data hypergraph; for example, two overlapping friend groups represented in your query hypergraph correspond to a single overall friend group in the data hypergraph. In this case, we could once again focus primarily on vertex alignment and introduce the non-exclusive overlap objective 
\begin{align*}
    R_{XO}(\sigma) &= \frac{1}{\lvert E_Q\rvert} \sum_{e \in E_Q} \max_{e^\prime \in E_D}\frac{\lvert \{\sigma(v) \mid v \in e\} \cap e^\prime \rvert}{\sqrt{\lvert e\rvert\lvert e^\prime\rvert} }.
\end{align*}

This itself corresponds to a more relaxed version of the incidence alignment problem where the injection constraint \(\lVert Y \rVert_1 \leq 1\) on the edge matching has been removed from Eq.\ \ref{eq:incidence}.

\subsection{Similarity Initialization}\label{sec:initialization}

For the purposes of similarity propagation, the computation of an initial similarity is important on multiple fronts. For one, it's what will be utilized to compute a sparse set of candidate pairs for large instances where considering all pairs for the final alignment would be computationally inefficient. What's more, a poor initial similarity may take a much longer time to converge (if ever) when running similarity propagation. As a result, it is very important to start with a good initial guess as to the similarity.

Previous works on similarity-based hypergraph alignments have utilized symmetric comparison functions of the form \(f(a,b)\) where \(f(a,b) \leq f(a,a)\) for all \(b\). This comparison function is then fed an attribute or vector of attributes associated with a given vertex in order to compute an initial similarity. For example, given vectors \(x,y\) one could set an initial similarity as \(S_{ij} = f(x_i, y_j).\)

For topology based graph alignment, degree alone has seen success as an input attribute for determining similarity \cite{qiuELRUNAEliminationRulebased2021,bigalign}. With this approach, one would use the attribute vectors \(x_v = \sum_{e \in E} B_{ve}\) and \(y_v = \sum_{e \in E} B_{ve},\) with an analogous definition using cardinality to compute the similarity between pairs of edges.
One natural option for an initial similarity for solving the incidence alignment problem in  Eq.\ \ref{eq:incidence} would be to instead use a weighted degree based on the edge normalized adjacency \(\tilde B.\)  Here, the attribute would be \(x_v = \sum_{e \in E} \tilde B_{ve}\) with an analogous definition for hyperedges.

However, the question remains as to how to incorporate the importance matrices \(W_V,W_E.\) To do so, we utilize a technique analogous to the idea of the eigenvector centrality for graphs. The eigenvector centrality of a graph \(G\) is given by the eigenvector corresponding to the largest eigenvalue 
\(\lambda_1\) of the adjacency matrix of \(G\) \cite{newman2008mathematics}. The eigenvector centrality of a vertex is equal to the value at its position in this eigenvector, and is measure of the importance of this vertex relative to the rest of the graph; furthermore, the eigenvector centrality is related to the degree of the given vertex as well as the value of the eigenvector centrality for its neighbors \cite{newman2008mathematics}. If scalability is a concern for certain applications, the eigenvector computation can be replaced by non-convergent but stabilized Gauss-Seidel relaxation in the spirit of algebraic distance \cite{chen2011algebraic,shaydulin2019relaxation}.

For this method, we utilize the left and right singular vectors \(\ell,r\) corresponding to the largest singular value \(\lambda\) of a given normalized incidence matrix \(\tilde B.\) Then \(\lambda \ell\) and \(\lambda r\) encode a notion of the importance of each vertex and each hyperedge respectively and can be used as attributes for initializing similarity. This can be extended to include the importance matrices \(W_V,W_E\) by embedding each of these matrices in to a single block and computing the left and right singular vectors of this block matrix
\begin{align}\label{eq:block_init}
    \argmax_{\substack{\ell_Q,r_Q,\ell_D,r_D \\ 
    \lVert r_Q\rVert_2^2 + \lVert r_D\rVert_2^2= \lVert \ell_Q\rVert_2^2 + \lVert \ell_D\rVert_2^2 = 1}} \begin{bmatrix}
        u^T_Q & r_D^T 
    \end{bmatrix}
    \begin{bmatrix}
        \frac{1}{\beta} W_V &  \tilde B_Q\\
        \tilde B_D^T & \frac{1}{\beta}W_E^T
    \end{bmatrix}\begin{bmatrix}
        \ell_D \\ r_Q 
    \end{bmatrix}.
\end{align}

 The vectors \(r_Q, \ell_Q, r_D, \ell_D\) then not only act as a representation of the importance of a vertex or hyperedge, but they are also influenced by which pairs are strongly encouraged by \(W_V,W_E\) i.e. if \((W_V)_{ij}\) is large, \((\ell_Q)_i\) and \((\ell_D)_j\) are also encouraged to be large.

A particularly useful option for \(W_V\) and \(W_E\) can be acquired by fixing part of the solution as \(X^*.\) Because the Frobenius inner product is bilinear, we get \begin{align}\label{eq:partial_solution}
\langle \tilde B_Q Y, (X + X^*)\tilde B_D \rangle_F &= \langle W_E, Y\rangle_F + \langle \tilde B_Q Y, X\tilde B_D \rangle_F, & W_E &= \tilde B_Q^TX^*B_D.
\end{align} 
Analogously, we can define \(W_V = \tilde B_QY^*B_D^T\) for some known matching between hyperedges. This allows for an iterative strategy, where if a complete match isn't found, the optimization may be run again with a fixed portion of the solution.

From here, we use the comparison function 
\(f(a,b) = \frac{\min(a,b)}{\max(a,b)},\)
as was used in the ELRUNA algorithm \cite{qiuELRUNAEliminationRulebased2021}. When a sparse initial similarity is needed, a KD-Tree can be used to efficiently get the top \(k\) most similar nodes \cite{chenCONEAlignConsistentNetwork2020}.



\subsection{Similarity Propagation}\label{sec:elimination_rules}

In this section, we present the two primary rules which we use to propagate similarities throughout the hypergraph, smoothing these values so that a pair with high similarity boosts the similarity of neighboring pairs. Likewise, if those neighboring pairs have a low similarity, this should instead decrease the similarity of the original pair. In this way, we can reshape the similarity matrix in order to better reflect the topological structure of the hypergraph.

\begin{algorithm}
    \caption{Propagate Similarities}\label{alg:propagate}
    \textbf{Input:} Similarity matrices \(X, Y;\) incidence matrices \(B_Q, B_D\)\\
    \For{$\_ \in \{1, \ldots, n_{iter}\}$}{
            \For{$(i,j) \in \texttt{nonzero(Y)}$}{
                Match \(N(e_i)\) to \(N(e_j)\), update \(Y_{ij}\) as in  Eq.\ \ref{eq:propagate}\;
            }
            \For{$(i,j) \in \texttt{nonzero(X)}$}{
                Match \(N(v_i)\) to \(N(v_j)\), update \(X_{ij}\) as in   Eq.\ \ref{eq:propagate}\;
            }
            Decay similarities as in Eq.\ \ref{eq:decay}\;
        }
\end{algorithm}

\subsubsection{Rule 1: Local Matching}

A naive implementation of similarity smoothing for hypergraphs punishes pairs of hyperedges which connect pairs of vertices that are not proportionally similar to one another. Instead, we attempt to find a local matching in the neighborhood of the pair of vertices whose similarities we're updating.

\begin{figure}
    \centering
    \setlength{\tabcolsep}{18pt}
    \begin{tabular}{cc}
        \includegraphics[width=0.35\linewidth]{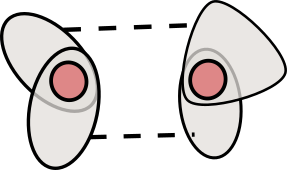} &  \includegraphics[width=0.35\linewidth]{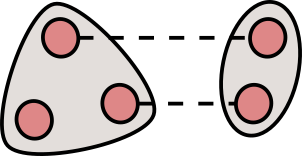} 
    \end{tabular}
    \caption{An update to the similarity for a pair of vertices is accomplished by aligning sets of hyperedges incident to those vertices, and analogously for the similarity of a pair of hyperedges. In this way, a given pair will have high similarity if and only if there is a matching with a high score in their neighborhood.}
    \label{fig:aligned}
\end{figure}

Finding a local matching prevents any one pair from contributing its similarity multiple times, preventing any one pair with high similarity from dominating all of its neighbors. Likewise, in the event that most pairs are dissimilar, it allows for the few pairs which are actually relevant to have a greater impact on the update. A depiction of the local matching is given in Fig. \ref{fig:aligned}.

Letting \(b_v\) and \(b_e\) be the greatest similarity that the vertex \(v\) or hyperedge \(e\) has with another node, the update is then 
\begin{align}\label{eq:propagate}
    X_{ij} = \frac{\texttt{dominant\_match}(Y, \tilde B_Q, \tilde B_D, i, j)}{\max\left(\sum_{u \in N(i)} b_u, \sum_{u \in N(j)} b_u\right)}, 
\end{align}
Where \(\texttt{dominant\_match}\) is the locally dominant matching algorithm described here \cite{mahantesh_multithread_matching}, acting on the submatrix of \(Y\) described by the i-th column of \(B_Q\) and the \(j\)-th column for \(B_D\). The update for \(Y\) is defined analogously.

\subsubsection{Rule 2: Similarity Cooling}

In order to encourage convergence, we also  include a cooling rule for the similarity propagation process. This rule penalizes pairs which are not locally dominant, pulling them down towards a lower threshold below which will be rounded to zero.

Let \(c_v = \frac{b_v}{T_v}\) and \(c_u = \frac{b_u}{T_u}.\) Then the cooling strategy utilized in ELRUHNA is defined by
\begin{align*}
    \text{decay}(s, t_1, t_2) &= \begin{cases}
        s & \max(t_1, t_2) \leq s\\
        s \sqrt{\frac{\min(t_1, t_2)}{\max(t_1, t_2)}} & \min(t_1, t_2) \leq s\\
        0 & \text{otherwise}
    \end{cases}
\end{align*}
\begin{align}\label{eq:decay}
    X^*_{ab} &= \text{decay}\left(X_{ab}^*,  \frac{b_a}{T_a},  \frac{b_b}{T_b}\right) & Y^*_{cd} &= \text{decay}\left(Y_{cd}^*,  \frac{b_c}{T_c},  \frac{b_d}{T_d}\right)
\end{align}
Pairs whose similarity falls between the dominant match for the two vertices will be scaled downwards at a rate proportional to the difference between the two values.

\section{Evaluation}

In this section, we will examine the performance of the ELRUHNA algorithm in comparison related methods such as BiG-Align, and Cone-Align \cite{bigalign,chenCONEAlignConsistentNetwork2020}. While HyperAlign \cite{doUnsupervisedAlignmentHypergraphs2024} is relevant, we are unable to compare against it at this time due to reproducibility issues. Likewise, methods such as HNN-HM are specialized towards an image processing domain and are not appropriate for the test cases available here.


We will focus on a number of instances from the HyperAlign paper ~\cite{doUnsupervisedAlignmentHypergraphs2024} and the XGI repositories of hypergraph datasets~\cite{xgi_repo}. This is a collection of hypergraphs from domains such as gene-disease association, co-authorship patterns, and social network analysis \cite{benson2018simplicial,pinero2020disgenet,bastian2009gephi}. For each hypergraph used in this paper, we follow the approach used in ~\cite{doUnsupervisedAlignmentHypergraphs2024} and compute the 2 core of the bipartite representation of the hypergraph; that is, we take the maximal induced subgraph such that every vertex has degree at least 2 and every hyperedge has a size of at least 2. This is because it is often infeasible to differentiate degree 1 vertices using only topological information.  The hypergraphs used range from several hundred vertices to tens of thousands after taking this core. Methods utilizing the bipartite representation and dense similarity matrices are constrained primarily by the maximum of \(\lvert V \rvert\) and \(\lvert E \rvert\): a summary of these sizes is presented in Table \ref{table:stats}. 

We focus primarily on aligning a given hypergraph with a noisy version of itself, which is a common downstream task to evaluate the alignment. In order to add noise to these hypergraphs, we utilize a probabilistic hypergraph model described in \cite{random-hypergraph}. For each random hyperedge, we assume that each vertex has a small independent probability \(p\) of being in that hyperedge. The size of this random hyperedge is then provided by a binomial distribution with \(\lvert V\rvert\) events and probability \(p\). We generally assume we are working with hypergraphs whose rank and maximum degree are significantly less than \(\lvert V \rvert\). As such, \(p << \lvert V\rvert\), and the size of each random hyperedge is approximately Poisson distributed with rate \(\lambda = \lvert V \rvert p.\) As such, the size of each hyperedge is determined by sampling from a Poisson distribution, then sampling that many vertices without replacement from \(V\). Throughout this paper, we choose \(\lambda = \bar{k},\) where \(\bar{k}\) is the average size of hyperedges in the hypergraph.

\begin{table} 
    \caption{ Hypergraph instances used in this paper, together with the size of their bipartite representation \(\lvert V \rvert + \lvert E \rvert\) and the average size of hyperedges in the sample.}
    \label{table:stats}
    \centering
    \renewcommand*{\arraystretch}{1.1}
    \begin{tabular}{cccc}
    \multicolumn{4}{c}{Small bipartite representation} \\ \hline
    & \(\lvert V \rvert\) &  \(\lvert E \rvert\) & \(\bar{k}\) \\ \hline
    email-Enron & 143 & 1457 & 3.1\\
    NDC-classes &  561 & 835 & 6.9\\
    diseasome & 114 & 152 & 2.5\\
    house-committees & 292 & 124 & 9.5\\
    senate-committees & 263 & 283 & 16.8
    \end{tabular}
    \qquad
    \begin{tabular}{cccc}
         \multicolumn{4}{c}{Large bipartite representation} \\ \hline
         & \(\lvert V \rvert\) & \(\lvert E \rvert\) & \(\bar{k}\) \\ \hline
coauth-Geology &  9537 & 36659 & 3.1 \\
contact-high-school & 327 & 4858 & 2.3 \\
coauth-History & 14968 & 20160 & 2.4 \\
contact-primary-school & 242 & 12704 & 2.4 \\
email-Eu & 900 & 24319 & 3.5 \\
threads-ask-ubuntu & 7522 & 31567 & 2.3 \\
    \end{tabular}
    \hfill
\end{table}


\subsection{Qualitative Assessment}

Let us compare the quality of ELRUHNA with competing methods on a number of small hypergraph examples which are suitable for dense methods. In particular, we compare against BiG-Align, which similarly works on the bipartite representation of the hypergraph structure. Additionally, we compare these results to that of Cone-Align, which is a state of the art method acting on the clique expansion of the hypergraphs.

\begin{figure}[htbp]
    \centering
    \setlength{\tabcolsep}{0pt} 
    \renewcommand{\arraystretch}{0} 
    \begin{tabular}{ccc}
        \includegraphics[width=0.32\textwidth]{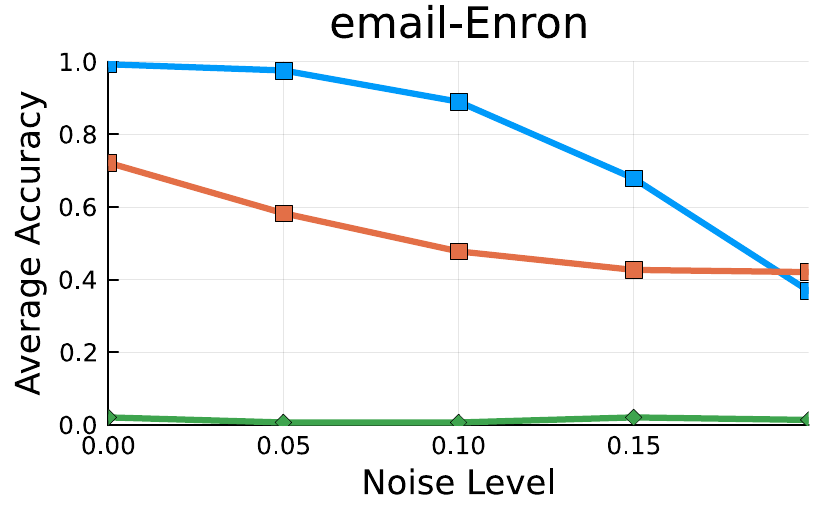} &
        \includegraphics[width=0.32\textwidth]{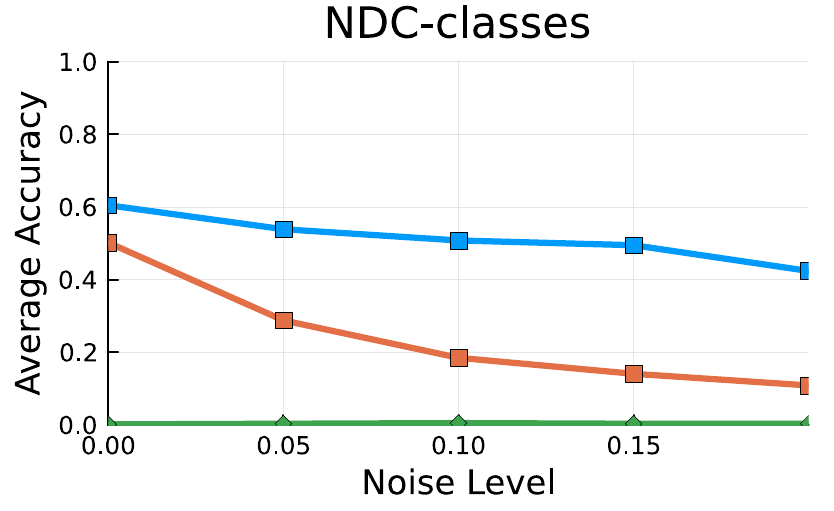} &
        \includegraphics[width=0.32\textwidth]{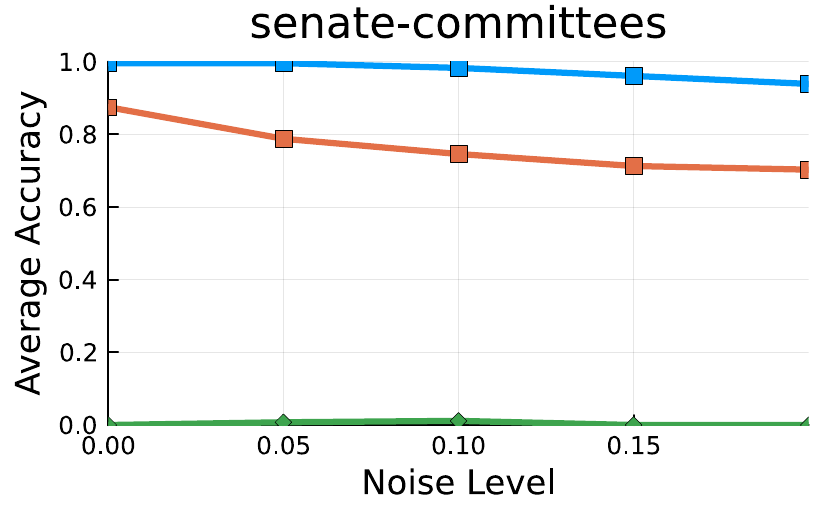} \\
        \includegraphics[width=0.32\textwidth]{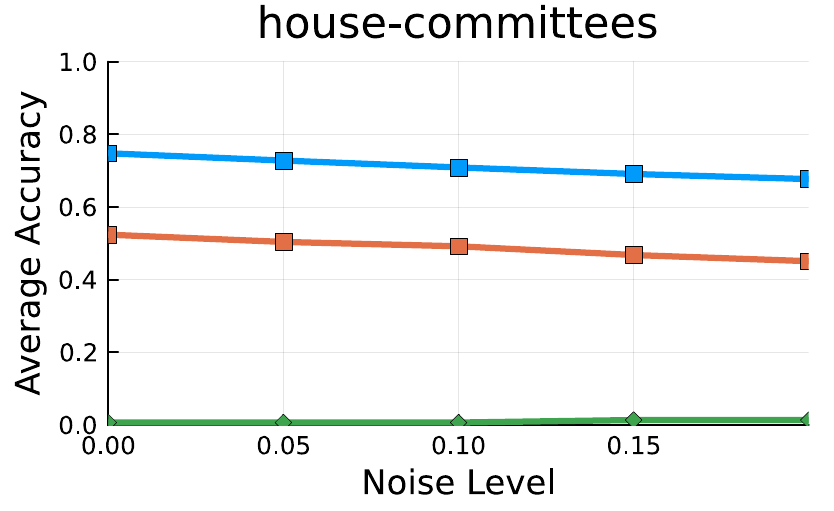} &
        \includegraphics[width=0.32\textwidth]{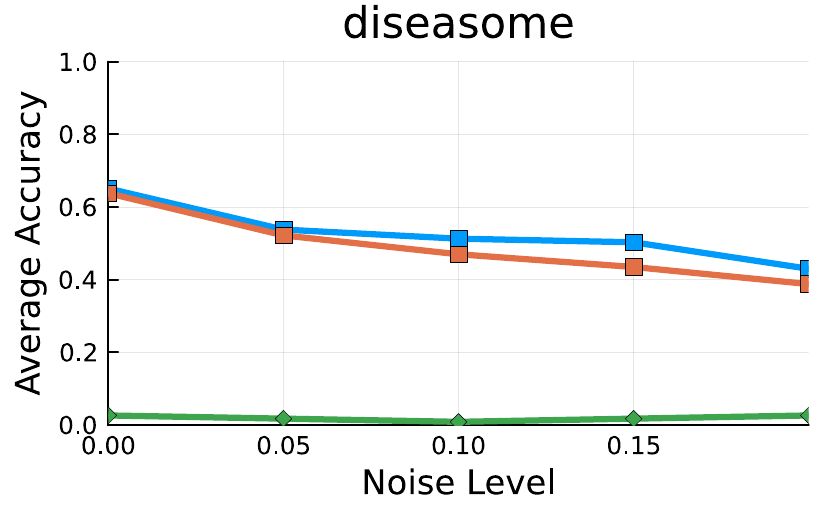} &
        \raisebox{20pt}{\includegraphics[width=0.15\textwidth]{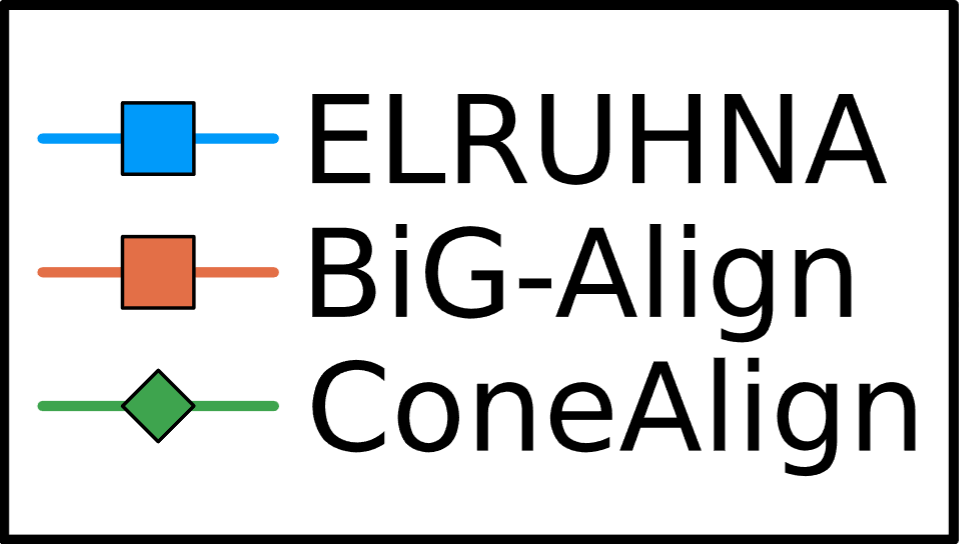}}
    \end{tabular}
    \caption{A comparison of ELRUHNA, BiG-Align, and Cone-Align for a collection of small hypergraph samples. Each subfigure gives the accuracy of the returned permutation as a function of the noise level in the input.}
    \label{fig:small_graph_tests}
\end{figure}

In particular, for these tests we compare the ground truth accuracy \(R_{acc}\) achieved by these methods on the provided hypergraphs. Each hypergraph is evaluated at 5 different noise levels across 10 different random seeds.  We summarize the average of these trials in Fig. \ref{fig:small_graph_tests}. This figure shows the average true accuracy over 10 runs as a function of the noise level of the input graph. On this collection, ELRUHNA was able to achieve over 90\% true accuracy on average for some instances in a low noise setting with a standard deviation of only \(0.01.\) This is as much as a 25\% increase in true accuracy over BiG-Align. These results significantly outperformed the results found using the clique expansion of the hypergraph, which struggled even in low noise setting. However, while ELRUHNA fairly consistently performs better on average than the other methods tested, we did note a significant increase in standard deviation at higher noise levels, up to 0.3. for some instances. Addressing these instabilities in the presence of large amounts of noise is a topic of future work. 

\begin{figure}
    \centering
    \includegraphics[width=.7\textwidth]{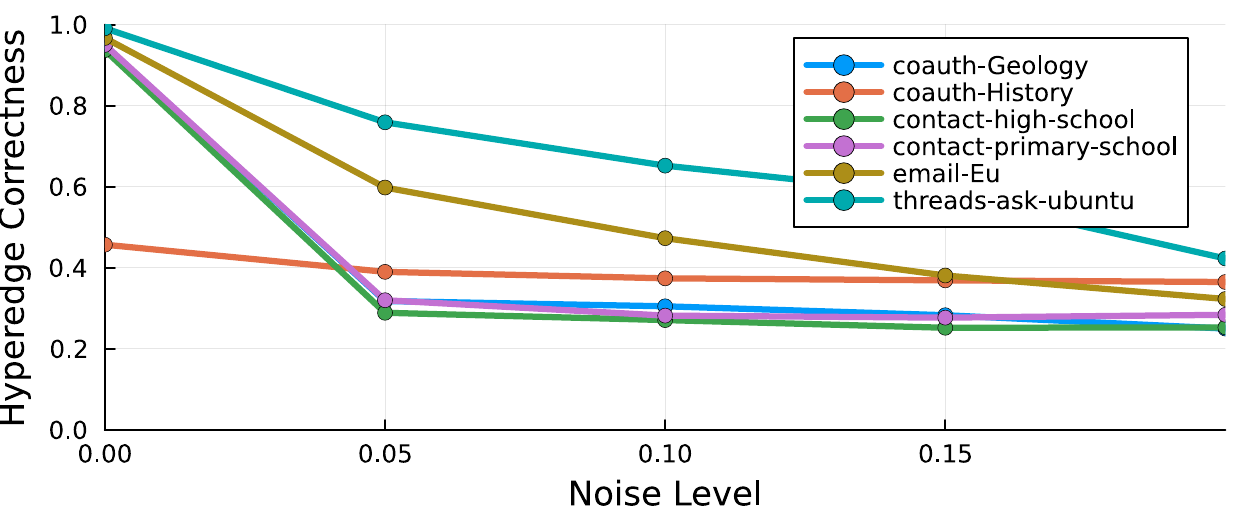}
    \caption{The hyperedge correctness found for a collection of larger larger hypergraph inputs as a function of noise level using the sparse mode of ELRUHNA.} 
    \label{fig:large_graph_tests}
\end{figure}

Next, we examine the behavior of ELRUHNA in a sparse setting, where it is intractable to represent the whole similarity matrix in dense form. The experiment setup remains generally the same as that used with the smaller hypergraph instances, although in this case we will be considering hyperedge correctness as a metric over true accuracy. when sparsifying, we consider only the top \(\lceil \log_2(\max(\lvert V\rvert, \lvert E\rvert))\rceil\) pairs. The results of these tests are shown in Fig. \ref{fig:large_graph_tests}. Notably, for the noise free setting where the input hypergraph is only permuted, ELRUHNA is still able to achieve a near perfect solution for some instances despite the aggressive sparsification. In the presence of noise, ELRUHNA fairly consistently outputs results in the range of \(15-25\%\) hyperedge correctness.

\section{Conclusion}

Hypergraph Alignment is an important NP-hard combinatorial optimization problem with applications in a wide variety of fields from biomedical to social sciences. In this paper, we have discussed the hyperedge overlap formulation of the hypergraph alignment problem in depth and provided an optimization formulation which connects this concept back to existing literature on alignment for bipartite graphs. What's more, we've provided an iterative framework for propagating similarities using the bipartite representation of the hypergraph using a pair of elimination rules. We've also discussed how to initialize these similarities by contrasting the importance of each pair using an extension of the eigenvector centrality to incidence matrices. Finally, we've explained how this can be extended to a sparse domain by iteratively updating the importance matrix \(W\) based on the current solution in order to guide a sparse local solution to a global one. We demonstrate an improvement of up to 25\% in terms of true accuracy compared to relevant alignment methods and show that we can acquire reasonable solutions for hypergraphs with tens of thousands of vertices in the bipartite representation. \emph{The source code and data are available at [link will be added upon acceptance]}.

\bibliographystyle{splncs_srt} 
\bibliography{align}

\begin{thebibliography}{10}

\bibitem{aladaug2013spinal}
Alada{\u{g}}, A.E., Erten, C.:
\newblock Spinal: scalable protein interaction network alignment.
\newblock Bioinformatics \textbf{29}(7) (2013)  917--924

\bibitem{random-hypergraph}
Barthelemy, M.:
\newblock Class of models for random hypergraphs.
\newblock Phys. Rev. E \textbf{106} (Dec 2022)  064310

\bibitem{bastian2009gephi}
Bastian, M., Heymann, S., Jacomy, M.:
\newblock Gephi: an open source software for exploring and manipulating networks.
\newblock In: Proceedings of the international AAAI conference on web and social media. Volume~3. (2009)  361--362

\bibitem{AlgorithmsLargeSparse}
Bayati, M., Gerritsen, M., Gleich, D.F., Saberi, A., Wang, Y.:
\newblock Algorithms for large, sparse network alignment problems.
\newblock In: 2009 Ninth IEEE International Conference on Data Mining, IEEE (2009)  705--710

\bibitem{bayati_sparse}
Bayati, M., Gleich, D.F., Saberi, A., Wang, Y.:
\newblock Message-passing algorithms for sparse network alignment.
\newblock ACM Trans. Knowl. Discov. Data \textbf{7}(1) (March 2013)

\bibitem{benson2018simplicial}
Benson, A.R., Abebe, R., Schaub, M.T., Jadbabaie, A., Kleinberg, J.:
\newblock Simplicial closure and higher-order link prediction.
\newblock Proceedings of the National Academy of Sciences \textbf{115}(48) (2018)  E11221--E11230

\bibitem{chen2011algebraic}
Chen, J., Safro, I.:
\newblock Algebraic distance on graphs.
\newblock SIAM Journal on Scientific Computing \textbf{33}(6) (2011)  3468--3490

\bibitem{chenCONEAlignConsistentNetwork2020}
Chen, X., Heimann, M., Vahedian, F., Koutra, D.:
\newblock {{CONE-Align}}: {{Consistent Network Alignment}} with {{Proximity-Preserving Node Embedding}}.
\newblock In: Proceedings of the 29th {{ACM International Conference}} on {{Information}} \& {{Knowledge Management}}. (October 2020)  1985--1988

\bibitem{doUnsupervisedAlignmentHypergraphs2024}
Do, M.T., Shin, K.:
\newblock Unsupervised alignment of hypergraphs with different scales.
\newblock In: Proceedings of the 30th ACM SIGKDD Conference on Knowledge Discovery and Data Mining. (2024)  609--620

\bibitem{friedlandSpecialIssuePolynomial2022a}
Friedland, S., Lasserre, J.B., Lim, L.H., Nie, J.:
\newblock Special {{Issue}}: {{Polynomial}} and {{Tensor Optimization}}.
\newblock Mathematical Programming \textbf{193}(2) (June 2022)  511--512

\bibitem{mahantesh_multithread_matching}
Khan, A.M., Gleich, D.F., Pothen, A., Halappanavar, M.:
\newblock A multithreaded algorithm for network alignment via approximate matching.
\newblock In: SC '12: Proceedings of the International Conference on High Performance Computing, Networking, Storage and Analysis. (2012)  1--11

\bibitem{bigalign}
Koutra, D., Tong, H., Lubensky, D.:
\newblock Big-align: Fast bipartite graph alignment.
\newblock In: 2013 IEEE 13th international conference on data mining, IEEE (2013)  389--398

\bibitem{xgi_repo}
Landry, N.W., Lucas, M., Iacopini, I., Petri, G., Schwarze, A., Patania, A., Torres, L.:
\newblock Xgi: A python package for higher-order interaction networks.
\newblock Journal of Open Source Software \textbf{8}(85) (2023)  5162

\bibitem{HNNHM}
Liao, X., Xu, Y., Ling, H.:
\newblock Hypergraph neural networks for hypergraph matching.
\newblock In: 2021 IEEE/CVF International Conference on Computer Vision (ICCV). (2021)  1246--1255

\bibitem{liaoHypergraphNeuralNetworks2021}
Liao, X., Xu, Y., Ling, H.:
\newblock Hypergraph {{Neural Networks}} for {{Hypergraph Matching}}.
\newblock In: Proceedings of the {{IEEE}}/{{CVF International Conference}} on {{Computer Vision}}. (2021)  1266--1275

\bibitem{mao-etal-2021-alignment}
Mao, X., Wang, W., Wu, Y., Lan, M.:
\newblock From alignment to assignment: Frustratingly simple unsupervised entity alignment.
\newblock In Moens, M.F., Huang, X., Specia, L., Yih, S.W.t., eds.: Proceedings of the 2021 Conference on Empirical Methods in Natural Language Processing, Online and Punta Cana, Dominican Republic, Association for Computational Linguistics (November 2021)  2843--2853

\bibitem{newman2008mathematics}
Newman, M.E.:
\newblock The mathematics of networks.
\newblock The new palgrave encyclopedia of economics \textbf{2}(2008) (2008)  1--12

\bibitem{neyshabur2013netal}
Neyshabur, B., Khadem, A., Hashemifar, S., Arab, S.S.:
\newblock Netal: a new graph-based method for global alignment of protein--protein interaction networks.
\newblock Bioinformatics \textbf{29}(13) (2013)  1654--1662

\bibitem{pinero2020disgenet}
Pi{\~n}ero, J., Ram{\'\i}rez-Anguita, J.M., Sa{\"u}ch-Pitarch, J., Ronzano, F., Centeno, E., Sanz, F., Furlong, L.I.:
\newblock The disgenet knowledge platform for disease genomics: 2019 update.
\newblock Nucleic acids research \textbf{48}(D1) (2020)  D845--D855

\bibitem{qiuELRUNAEliminationRulebased2021}
Qiu, Z., Shaydulin, R., Liu, X., Alexeev, Y., Henry, C.S., Safro, I.:
\newblock {{ELRUNA}}: {{Elimination Rule-based Network Alignment}}.
\newblock ACM J. Exp. Algorithmics \textbf{26} (December 2021)  1--32

\bibitem{shaydulin2019relaxation}
Shaydulin, R., Chen, J., Safro, I.:
\newblock Relaxation-based coarsening for multilevel hypergraph partitioning.
\newblock Multiscale Modeling \& Simulation \textbf{17}(1) (2019)  482--506

\bibitem{user_correspondence}
Tan, S., Guan, Z., Cai, D., Qin, X., Bu, J., Chen, C.:
\newblock Mapping users across networks by manifold alignment on hypergraph.
\newblock In: Proceedings of the Twenty-Eighth AAAI Conference on Artificial Intelligence. AAAI'14, AAAI Press (2014)  159–165

\bibitem{xiangCuAlignScalableNetwork2023}
Xiang, L., Khan, A., Ferdous, S.M., Aravind, S., Halappanavar, M.:
\newblock {{cuAlign}}: {{Scalable Network Alignment}} on {{GPU Accelerators}}.
\newblock In: Proceedings of the {{SC}} '23 {{Workshops}} of {{The International Conference}} on {{High Performance Computing}}, {{Network}}, {{Storage}}, and {{Analysis}}. {{SC-W}} '23, New York, NY, USA, Association for Computing Machinery (November 2023)  747--755

\bibitem{yanAdaptiveDiscreteHypergraph2018}
Yan, J., Li, C., Li, Y., Cao, G.:
\newblock Adaptive {{Discrete Hypergraph Matching}}.
\newblock IEEE Transactions on Cybernetics \textbf{48}(2) (February 2018)  765--779

\bibitem{ShortSurveyRecent}
Yan, J., Yin, X.C., Lin, W., Deng, C., Zha, H., Yang, X.:
\newblock A short survey of recent advances in graph matching.
\newblock In: Proceedings of the 2016 ACM on international conference on multimedia retrieval. (2016)  167--174

\end{thebibliography}











\end{document}